\documentstyle[psfig,referee]{laa}
\thesaurus{03.13.1; 08.19.1; 11.19.4; 11.03.1; 11.09.2; 12.12.1}
\title{The statistics of the gravitational field arising from an
inhomogenous system of particles}
\author{A. ~Del Popolo\inst{1,2}, M. Gambera\inst{1,3} }
\offprints{A. ~Del Popolo, M. ~Gambera}
\institute{ $^1$ Istituto di Astronomia dell'Universit\`a di Catania, 
Citt\`a Universitaria, Viale A.Doria, 6 - I 95125 Catania, Italy\\
$^2$ Dipartimento di Matematica, Universit\`{a} Statale di Bergamo, Piazza Rosate, 2 - I 24129 Bergamo, ITALY\\
$^3$ Osservatorio Astrofisico di Catania and CNR-GNA, 
Viale A.Doria, 6 - I 95125 Catania, ITALY \\}
\date{}
\begin{document}
\maketitle
\begin{abstract}
In this paper we extend Chandrasekhar and von Neumann's analysis of the
statistics of the gravitational field to systems in which particles 
({\it e.g.} stars, galaxies) are not homogeneously distributed.  
We derive a
distribution function $ W({\bf F},d{\bf F}/dt)$ giving the joint
probability that a test
particle is subject to a force {\bf F} and an associated rate of change of 
{\bf F} given by d{\bf F}/dt. We calculate the first moment of d{\bf F}/dt 
to study the effects of inhomogenity on dynamical friction. 
\end{abstract}

\section{Introduction}
					    
The study of the statistics of the fluctuating gravitional force in infinite 
homogeneous systems was pionered by Chandrasekhar \& von
Neumann in two classical papers (Chandrasekhar \& von Neumann
1942, 1943 hereafter CN43) and in several other papers by Chandrasekhar 
(1941, 1943a,b,c,d,e, 1944a and 1944b). The analysis of
the fluctuating gravitional field, developed by the quoted authors, was
formulated by means of a statistical treatment. In their papers 
Chandrasekhar \& von Neumann considered a system in which the
stars are distributed according to a uniform probability density, 
no correlation among the positions of the stars is present and 
where the number of stars constituting the system tends to infinity while
keeping the density constant.
The force {\bf F}, for
unit mass, acting on a star of a generical star system is given by the
well known equation:
\begin{equation}
{\bf F} = - G  \sum_{i=1}^N \frac{M_i}{|{\bf r}_{i}|^{3}} {\bf r}_{i}
\label{eq:uno}
\end{equation}
where $ M_{}$ is the mass of the i-th field star,
$ N$ is the total number of
the stars in the system and $ {\bf r}_{i}$ is the 
position vector of the i-th test star relative to the field one.
The summation includes all the neighboring stars. The motion of the 
stars in the neighborhood of the test star produces a time variation
of {\bf F}.
The exact dependence of {\bf F} on the position and time  
cannot be exactly predicted,
while it is possible to study the fluctuation of {\bf F} from a 
statistical point of view.\\
Two distributions are fundamental for the description of the fluctuating 
gravitional field: 
\begin{itemize}
\item 1) $ W({\bf F})$ which gives the probability that a test star
is subject to a force ${\bf F}$ in the range ${\bf F}$, ${\bf F}$+
d${\bf F}$;
\item 2) $ W({\bf F},{\bf f})$ which gives the
joint probability that the star experiences a force {\bf F} and a rate 
of change {\bf f}, where $ {\bf f} = d{\bf F}/dt $.
\end{itemize}
The first distribution, known as Holtsmark's law (Holtsmark 1919), in the 
case of homogeneous distribution of the stars, gives information only 
on the number of stars experiencing a given force but it does not 
describe some fundamental features of the fluctuations in 
the gravitional field such as the {\it speed of the fluctuations} and 
the dynamical friction. These features can be 
described using the second distribution $ W({\bf F},{\bf f})$.
As shown by CN43, the
{\it speed of the fluctuations}
can be adequately expressed in terms of the mean life of a 
state {\bf F}:
\begin{equation}
T = \frac{ |{\bf F} |}{\sqrt{\langle | {\bf f}_{\bf F}^{2} | \rangle}}
\label{eq:du}
\end{equation}
where $ \langle | {\bf f}^2 | \rangle$ is the second moment of {\bf f}.
Hence, for the definition of the speed of fluctuations and of the 
dynamical friction one must determine the distribution $ W({\bf F},{\bf f})$. 
For a test star moving whith velocity {\bf v} in
a sea of field stars characterized by a random probability distribution 
of the velocities, $ \Phi({\bf u})$, we may write:
\begin{equation}
\langle {\bf V} \rangle = \langle {\bf u} \rangle - {\bf v} = - {\bf v}
\label{eq:tr}
\end{equation}
where ${\bf V}$ represents the velocity of a typical field star relative
to the one under consideration, ${\bf u}$ denotes the velocity of a
field star. 
This asymmetry of the distribution of the relative velocities produces, as
shown by CN43, a
deceleration of the test star in the direction of motion.
This effect is known, after Chandrasekhar papers, as "{\it dynamical friction}".
Some information on dynamical friction can be obtained by means of the 
first moment of {\bf f}. As shown by CN43:
\begin{equation}
\langle \frac{d{\bf F}}{dt} \rangle_{{\bf F},{\bf v}} = \frac{-2 \pi}{3} G m n  B(\beta) \left[ {\bf v} - \frac{3 {\bf F} \cdot {\bf v}}{ | {\bf F} |^2} \cdot {\bf F} \right]
\label{eq:qu}
\end{equation}
\begin{equation}
\langle \frac{d | {\bf F} |}{dt} \rangle_{{\bf F},{\bf v}} = \frac{ 4 \pi}{3} G m n  B(\beta) \frac{{\bf F} \cdot {\bf v}}{ | {\bf F} |}
\label{eq:cin}
\end{equation}
where $ m$ is the mass of a field star, $ n_{l}$ is the local density, and 
$ B(\beta)$ is definited in CN43 Eq. (98).
These equations show that the amount of acceleration in the direction 
of $- {\bf v}$, when $ {\bf v} \cdot {\bf F} \le 0$, is greater than that 
in the direction $+ {\bf v}$, when $ {\bf v} \cdot {\bf F} \ge 0$. 
The star suffers a deceleration being the a priori probability that 
$ {\bf v} \cdot {\bf F} \ge 0$ equal to the probability that 
$ {\bf v} \cdot {\bf F} \le 0$. \\
Several authors have stressed
the importance of stochastic forces and in particular dynamical friction
in determining the observed
properties of clusters of galaxies (White 1976; Kashlinsky 1986, 1987)
while others studied the role of dynamical friction in the orbit decaying
of a satellite moving around a galaxy or in the merging
scenario (Bontekoe \& van Albada 1987; Seguin \& Dupraz 1996;
Dominguez-Tenreiro \& Gomez-Flechoso 1998) which is not only
the framework for galaxy
formation picture in hierarchical cosmological models, but also
important for the study of
particular aspects of the evolution of a number of astronomical
systems, such as galactic nuclei, cD galaxies in rich galaxy clusters.
Finally, here, we want to remember as
other the statistical description of dynamical friction 
other works have been made on it based on different approaches 
for example:
Fokker Planck equation based polarization cloud (Rosenblunth et al.  1957; 
Binney \& Tremaine 1987); resonant particle interactions (Tremaine \& 
Weinberg 1984; Weinberg 1986); fluctation dissipation (Berkenstein \& Maoz 
1992; Maoz 1993).\\
Chandrasekhar's theory (and in particular his
classical formula - see Chandrasekhar 1943b)
is widely employed to quantify dynamical friction
in a variety of situations, even if
all the theory developed by the quoted authors is based on the
hypothesis that the stars are distributed uniformly and 
it is well known that in stellar systems, the stars are not 
uniformly distributed,
(Elson et al. 1987; Wybo \& Dejonghe 1996; Zwart et al. 1997) 
as well as in galactic systems, the galaxies are not uniformly distributed 
(Peebles 1980; Bahcall \& Soneira 1983; Sarazin 1988; Liddle, \& Lyth 1993; 
White et al. 1993; Strauss \& Willick 1995).
It is evident
that an analysis of dynamical friction taking account of the
inhomogeneity of astronomical systems can provide a more realistic
representation of the evolution of these systems itself.
Moreover from a pure theorethical ground we expect that inhomogeneity affects all the aspects of the fluctuating 
gravitational field (Antonuccio \& Colafrancesco 1994; Del Popolo 1994; 
Del Popolo et al. 1996; Del Popolo \& Gambera 1996, 1997; Gambera 1997). 
Firstly the Holtsmark distribution is no more correct 
for inhomogeneous systems. For these systems, as shown by Kandrup 
(1980a,b, 1983), the Holtsmark distribution must be substituted 
with a generalized form of the Holtsmark distribution
characterized by a shift of $W({\bf F})$
towards larger forces when inhomogeneity increases. This result 
was already suggested by the numerical simulations of Ahmad \& Cohen 
(1973, 1974). 
Hence when the inhomogeneity increases
the probability that a test particle experiences a
large force increases, secondly 
$ W({\bf F},{\bf f}) $ is changed by inhomogeneity. 
Consequently, the values of the mean life of a state, 
the first moment of $ {\bf f}$ and the dynamical friction force
are changed by inhomogeneity with respect to those of homogeneous systems.\\
This paper must be intended as the first part of a work
pointed to:
\begin{itemize}
\item a) study the effects of inhomogeneity on the distribution functions
of the stochastic forces and on dynamical friction (present paper);
\item b) test the result against N-body simulations; 
\item c) find a formula that describes dynamical friction in homogeneous and 
inhomogeneous systems only on the basis of
the statistical theory.
\end{itemize}
Before continuing we want
to stress that when we speak of inhomogeneity we refer to inhomogeneity
in position distribution and not to that of velocity distribution. Our
work  follows the spirit of Kandrup's (1980a) in the sense that
we are interested in
the effect of a non-uniform distribution in position of stars on the
distributions of the stochastic force.\\
The plan of the paper is the following: in Sect. ~2 we sketch the
calculations needed to obtain the
distribution function $ W({\bf F},{\bf f}) $ after having released 
the hypothesis of homogeneity.
The complete calculations are developed in the appendix.
Then in Sect. ~3 we calculate the first
moment of {\bf f} and in Sect. ~ 4 we 
show how dynamical friction is influenced by inhomogeneity. 
Finally, in Sect. 5 we draw our conclusions.

\section{The distribution function $ W({\bf F},{\bf f})$ in inhomogeneous systems} 

To calculate $ W({\bf F},{\bf f})$ in an 
inhomogeneous system we consider a particle moving with a velocity ${\bf v}$, 
subject to a force, per unit mass, given by Eq. (\ref{eq:uno}) 
and to a rate of change given by
\begin{equation}
{\bf f} = \frac{d{\bf F}}{dt} = G \sum_{i=1}^N M_i \left[ \frac{{\bf V}_i}{|{\bf r}_{i}|^{3}} - \frac{3 {\bf r}_{i} ({\bf r}_{i}{\bf V}_i)}{|{\bf r}_{i}|^{5}} \right]
\label{eq:sei}
\end{equation}
where $ {\bf V}_{i}$ is the velocity of the field particle relative to 
the test one. \\
The expression of $ W({\bf F},{\bf f})$ is given following Markoff's
method by (CN43):
\begin{eqnarray}
W({\bf F},{\bf f}) & = & \frac{1}{64 \pi^{6}} \int_{0}^{\infty}
\int_{0}^{\infty}  A({\bf k},{\bf \Sigma}) \cdot  \nonumber \\
                   &   &\left\{ \exp{[-i({\bf k}{\bf \Phi} +
{\bf \Sigma}{\bf \Psi})]} \right\} d{\bf k} d{\bf \Sigma}
\label{eq:set}
\end{eqnarray}
with $ A({\bf k}, {\bf \Sigma})$ given by 
\begin{equation}
A({\bf k}, {\bf \Sigma}) = e^{-nC({\bf k}, {\bf \Sigma})} \label{eq:setbis}
\end{equation}
being 
\begin{equation}
C({\bf k}, {\bf \Sigma}) = \int_{0}^{\infty}
\int_{- \infty}^{\infty}
\int_{- \infty}^{\infty} \tau \left[ 1 - \exp{i({\bf k}{\bf \Phi}
+ {\bf \Sigma}{\bf \Psi})} \right] d{\bf r} d{\bf V} dM
\label{eq:ot}
\end{equation}
where $ n$ is the average number of stars per unit volume while $ {\bf \Phi}$ 
and $ {\bf \Psi}$ are given by the following relations: 
\begin{equation} 
{\bf \Phi} = G \frac{M}{|{\bf r} |^{3}} {\bf r} 
\label{eq:nov}
\end{equation}
\begin{equation} 
{\bf \Psi} = \frac{d{\bf F}}{dt} = M \left[ \frac{{\bf V}}{|{\bf r} |^{3}} - \frac{3 {\bf r} ({\bf r} {\bf V})}{|{\bf r} |^{5}} \right]
\label{eq:die}
\end{equation}
and $ \tau ({\bf V},{\bf r},M) d {\bf V} d {\bf r}dM $ is
the probability that a star has velocity in the
range $ {\bf V}, {\bf V} + d{\bf V}$, positions in $ {\bf r}, {\bf r} + d{\bf r}$
and mass in $ M, M + dM$.\\
Now we suppose that 
$\tau$ is given by:
\begin{equation}
\tau = \frac{a}{r^p} \psi (j^2(M) |{\bf u}|^2)
\label{eq:un}
\end{equation}
where $ a$ is a constant
that can be obtained from the normalization condition for  $\tau$, $j$ a
parameter (of dimensions of velocity$^{-1}$), $\psi$ an arbitrary
function, $\bf u$ the velocity of a field star.
In other words we assume, according to CN43 and Chandrasekhar
\& von Newmann (1942), that the distribution of velocities is spherical,
i.e. the distribution function is 
$\psi({\bf u}) \equiv  \psi (j^2(M) |{\bf u}|^2) $,
but differently from the quoted papers we suppose that the positions
are not equally likely for stars, that is the stars are
inhomogeneously distributed in space.
A lenghty calculation leads us (see Appendix for a derivation and the
meaning of simbols) to find the function
$A({\bf k}, {\bf \Sigma})$:
\begin{eqnarray}
A({\bf k}, {\bf \Sigma}) & = & e^{  -\tilde{a} k^{\frac{3-p}{2}}} \{1- \; i g p({\bf k}, {\bf \Sigma})   \nonumber \\
                            &   & + \; \tilde{b} k^{\frac{- (3 
+ p)}{2}} ] \cdot  [ Q({\bf \Sigma}) + k R({\bf \Sigma}) ] \} 
\label{eq:trense}
\end{eqnarray} 
This last equation introduced into Eq. (\ref{eq:set}) solves the problem
of finding the distribution $W(\bf F,\bf f)$ and makes it possible to find
the moments of $\bf f$ that give information regarding the dynamical
friction.

\section{Evalutation of $ \overline f$}

As we stressed in the introduction, the study of the dynamical friction 
is possible when we know the first moment of $ {\bf f}$. This calculation 
can be done using the components of $ {\bf f}$ ($ f_i, f_j, f_k$) in the 
system of coordinates previously introduced. We have that:
\begin{equation}
f_i =  \frac{\int_{- \infty}^{\infty} W({\bf F},{\bf f})  f_i d{\bf f}}{W({\bf F})}
\label{eq:trenset}
\end{equation} 
and similar equations for the other components of the force. The distribution 
function $ W({\bf F})$, giving the number of stars subject to a force 
$ {\bf F}$, can be calculated as follows:
\begin{eqnarray}
W({\bf F}) & = & \frac{1}{64 \pi^{6}} \int_{0}^{\infty} \int_{0}^{\infty} \int_{0}^{\infty} \{ e^{[-i({\bf k}{\bf F} + {\bf \Sigma}{\bf f})]} \}  \nonumber \\
           &   &\cdot A({\bf k},{\bf \Sigma}) \; d {\bf k} d {\bf \Sigma} 
d {\bf f}
\label{eq:trenot}
\end{eqnarray}
integrating we find:
\begin{eqnarray}
W({\bf F}) =  \frac{1}{2 \pi^{2} F} \int_{0}^{\infty} \{ e^{ [ \; - a^2 k^{(3-p)/2} ] } \} \cdot  k \sin({k F}) dk
\label{eq:trenno}
\end{eqnarray}
This equation gives the generalized Holtsmark distribution obtained by 
Kandrup (1980a - Eq. 4.17) and provides the probability that a star is 
subject to a force $ {\bf F}$ in a inhomogeneous system.\\
As previously stressed, to calculate the first moment of $ {\bf f}$
we need only an approximated form for $ A({\bf k}, {\bf \Sigma})$:
\begin{equation}
A({\bf k}, {\bf \Sigma}) = \{ e^{[ \; - a^2 k^{(3-p)/2}]} \} \cdot [1 - i g p({\bf k}, {\bf \Sigma}) ] 
\label{eq:quar}
\end{equation}
Using this last expression for $A({\bf k}, {\bf \Sigma})$ and
Eq. (\ref{eq:cas}), Eq. (\ref{eq:cass}), Eq. (\ref{eq:trenset}), Eq. (\ref{eq:trenno}),
Eq. (\ref{eq:quar}) and performing a calculation similar to that by 
CN43 
the first moment of  $ {\bf f}$ is given by:
\begin{eqnarray}
\overline {\bf f} & = & -\left( \frac{1}{2} \right)^{\frac{3}{3 - p}} \cdot A(p) \cdot B(p)^{\frac{p}{3 - p}} \nonumber \\
                  &   &\cdot \frac {\alpha^{\frac{3}{3 - p}} G M L(\beta)}{ \pi H(\beta) \beta^{\frac{2 - p}{2}}} \cdot \left[ {\bf v} - \frac{3 {\bf F} \cdot {\bf v}}{ | 
{
\bf F} |^2} \cdot {\bf F} \right]
\label{eq:quaruno}
\end{eqnarray}
where
\begin{eqnarray}
L ( \beta) & = & 6 \int_{0}^{\infty} \left[
e^{(x/\beta)^{\frac{(3 - p)}{2}}} \right]
\left[ \frac{ \sin{x}}{x^{(2 - p)/2}} -
\frac{\cos{x}}{x^{p/2}} \right] dx \nonumber \\
           &   & - \; 2 \int_{0}^{\infty}
           \left[ e^{(x/\beta)^{\frac{(3 - p)}{2}}}
           \right] \cdot \frac{\sin{x}}{x^{(p - 2)/2}} dx
\label{eq:quardu}
\end{eqnarray}
and for $ p = 0$ Eq. (\ref{eq:quaruno}) reduces to:
\begin{equation}
\overline {\bf f} = - \frac{\alpha}{6} G m \left[ \frac{L(\beta)}{\pi \beta H(\beta)} \right]_{p = 0} \cdot \left[ {\bf v} - \frac{3 {\bf F} \cdot {\bf v}}{ | {\bf F} |^2} \cdot {\bf F} \right]
\label{eq:quartre}
\end{equation}
and consequently
\begin{eqnarray}
[ L ( \beta)]_{p = 0} & = & \int_{0}^{\infty}
e^{ \left[ - \; (x/\beta)^{3/2} \right] } \cdot \nonumber \\
                      &   &\left[ \frac{ 6 \sin{x}}{x} - 6 \cos{x} + 2 x \sin{x} \right] dx
\label{eq:quarqua}
\end{eqnarray}
this last expression can also be written as:
\begin{equation}
[ L ( \beta)]_{p = 0} = 3 \pi \int_{0}^{\beta} H( \beta) d \beta - \pi \beta H( \beta)
\label{eq:quarcin}
\end{equation}
being
\begin{equation}
H ( \beta) = \frac{2}{\pi \beta} \int_{0}^{\infty} \left\{ e^{ \left[ (\frac{ - \; x }{\beta})^{3/2}\right]}  \right\} \cdot  x \sin{x} dx
\label{eq:quarsei}
\end{equation}
In this way we can written Eq. (\ref{eq:quartre}) as:
\begin{equation}
\overline {\bf f} = - \frac{2 \pi}{3} G m n \left[ \frac{3 \cdot \int_{0}^{\beta} H(\beta) d \beta }{ \beta \cdot H(\beta)} \; - \; 1 \right] \cdot \left[ {\bf v} - \frac{3 {\bf F} \cdot {\bf v}}{ | {\bf F} |^2} \cdot {\bf F} \right]
\label{eq:quarset}
\end{equation}
this last equation coincides with Eq. (105) by CN43.\\
The results obtained by us for an inhomogeneus system 
are different [see Eq. (\ref{eq:quaruno})], as expected,
from that obtained by CN43 for
a homogeneous system (CN43 - Eq. 105). At the same time 
it is very interesting
to note that for $ p = 0$ (homogeneous system) our result coincides,
as obvious, with the
results obtained by CN43. In a inhomogeneous system, in a similar 
way to what happens in a homogeneus system, $ {\bf f}$ depends
on $ {\bf v}$, $ {\bf F}$ and $ \theta$ (the angle between $ {\bf v}$ and 
$ {\bf F}$) while differently from homogeneous systems, $ {\bf f}$ 
is a function of the inhomogeneity parameter
$p$. The dependence of $ {\bf f}$ on $p$ is not only due to the
functions $A(p)$, $ B(p)$ and to the density parameter $ \alpha$ but
also to the parameter $ \beta=|\bf F|/Q_{H}$. In fact in
inhomogeneous systems the {\it normal} field $Q_{H}$ is given by
$Q_{H}=GM(\alpha B(p)/2)^{2/(3-p)}$, clearly dependent on $p$.

\section{Dynamical friction in inhomogeneous systems}

The introduction of the notion of dynamical friction is due to CN43. In the 
stochastic formalism developed by CN43 the dynamical friction is discussed 
in terms of $ {\bf f}$:
\begin{equation}
\overline {\bf f} =  \frac{-2 \pi}{3} G m n  B(\beta) \left[ {\bf v} - \frac{3 {\bf F} \cdot {\bf v}}{ | {\bf F} |^2} \cdot {\bf F} \right]
\label{eq:quarot}
\end{equation}
where
\begin{equation}
B( \beta) =  \frac{3 \cdot \int_{0}^{\beta} W(\beta) d \beta }{ \beta \cdot W(\beta)} \; - \; 1
\label{eq:quarno}
\end{equation}
and $ \beta = | {\bf F}| / Q_{H} = | {\bf F}| / 2.603 G M n^{2/3}$.
As shown by CN43, the origin of dynamical friction is due to 
the asymmetry in the distribution of relative velocities. As previously 
told, if a test star moves with velocity $ {\bf v}$ in a spherical 
distribution of field stars, namely $ \phi( \bf u)$ then we have that:
\begin{equation}
\overline{\bf V} = \overline{{\bf u} - {\bf v}} = - {\bf v}
\label{eq:cinqqua}
\end{equation}
The asymmetry in the distribution of relative velocities is conserved 
in the final Eq. (\ref{eq:quarot}). In fact from Eq. (\ref{eq:quarot}) 
we have:
\begin{equation}
\frac{d |{\bf F}|}{dt} = \frac{4 \pi}{3} G M n B(\beta) \cdot \frac{{\bf v F}}{{\bf F}} \label{eq:cinqcin}
\end{equation}
(CN43). This means that when $ {\bf v} \cdot {\bf F} \ge 0$ then 
$ \frac{d |{\bf F}|}{dt} \ge 0$; 
while when $ {\bf v} \cdot {\bf F} \le 0$ then $ \frac{d |{\bf F}|}{dt} \le 0$.
As a consequence, when ${\bf F}$ has a positive component in the direction of
${\bf v}$,  ${|\bf F|}$ increases on average; while if  ${\bf F}$ has
a negative component in the direction of  ${\bf v}$,  ${|\bf F|}$
decreases on average. 
Moreover, the star suffers a greater amount of acceleration in the
direction $ - {\bf v}$ when
$ {\bf v} \cdot {\bf F} \le 0$ than in the direction $ +{\bf v}$ when 
$ {\bf v} \cdot {\bf F} \ge 0$.\\ 
In other
words the test star suffers, statistically, an equal number of 
accelerating and decelerating impulses. Being the modulus of 
deceleration larger than that of acceleration the star slows down.\\
At this point we may show how dynamical friction changes due to 
inhomogeneity. From Eq. (\ref{eq:quaruno}) we see that
$ \frac{d{\bf F}}{dt}$ differs from that obtained in homogeneus system 
only for the presence of a dependence on the inhomogeneity parameter 
$ p$. If we divide Eq. (\ref{eq:quaruno}) for Eq. (\ref{eq:quarot}) we 
obtain:
\begin{eqnarray}
\frac{ \left( \overline{\frac{d{\bf F}}{dt}} \right)_{Inh.}}
{ \left( \overline{\frac{d{\bf F}}{dt}} \right)_{Hom.}} 
& = & - \left( \frac{1}{2} \right)^{\frac{3}{3 - p}} \; B(p)^{\frac{p}{3 - p}} \; \frac{\alpha^{\frac{3}{3 - p}} L(\beta)}{ \pi H(\beta) \beta^{\frac{2 - p}{2}}} \cdot \frac{3 \cdot A(p)}{2 \pi n B(\beta)} \; \nonumber \\
& = & \; - \left( \frac{1}{2} \right)^{\frac{6 - p}{3 - p}} \cdot  \frac{3 \alpha^{\frac{3}{3 - p}} L(\beta) B(p)^{\frac{p}{3 - p}} \cdot A(p) }{n \cdot \pi^{2} H(\beta) B(\beta) \beta^{\frac{2 - p}{2}}}
\label{eq:cinqsei}
\end{eqnarray}
If we consider a homogeneous system, $ p=0$, the previous equation
reduces to:
\begin{equation}
\frac{ \left( \overline{\frac{d{\bf F}}{dt}} \right)_{Inh.}}
{ \left( \overline{\frac{d{\bf F}}{dt}} \right)_{Hom.}} 
= 1
\end{equation}
In the case of an inhomogeneous system, $ p \neq 0$, we see that:
\begin{equation}
\frac{ \left( \overline{\frac{d{\bf F}}{dt}} \right)_{Inh.}}
{ \left( \overline{\frac{d{\bf F}}{dt}} \right)_{Hom.}} 
=n^{p/(3-p)} F(\beta(n,p))
\end{equation}
where
\begin{equation}
F[\beta(n,p)] =  \; - \left( \frac{1}{2} \right)^{\frac{6 - p}{3 - p}} \cdot B(p)^{\frac{p}{3 - p}} \cdot \frac{3 L(\beta) \cdot A(p)}{n \cdot \pi^{2} H(\beta) B(\beta) \beta^{\frac{2 - p}{2}}}
\end{equation}
\begin{figure}
\psfig{file=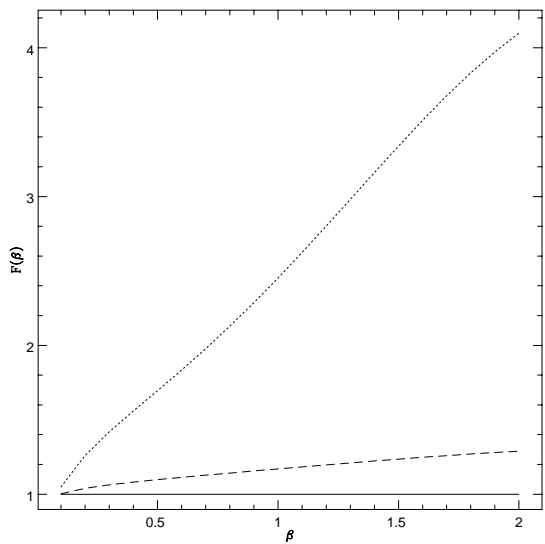,width=8cm}
\caption[]{The function $F(\beta)$ for several values of the inhomogeneity 
parameter $p$; solid line $p=0$, dashed line $p=0.1$, dotted line $p=0.5$}.
\end{figure}
As we show in Fig. 1, 
this last  equation is an increasing function of $p$. This means
that for increasing values of $ p$ 
the star suffers an even greater amount of acceleration in the
direction $ - {\bf v}$ when
$ {\bf v} \cdot {\bf F} \le 0$ than in the direction $ +{\bf v}$ when 
$ {\bf v} \cdot {\bf F} \ge 0$, with respect to the homogeneous case.
This is due to the fact that the difference
between the amplitude of the decelerating impulses and the accelerating ones 
is, as in homogeneous systems, statistically negative, but now larger,
being the scale factor greater.
This finally means that, for a given value of $n$, 
the dynamical friction increases with increasing inhomogeneity in
the space distribution of stars 
(it is interesting to note that this effect is fundamentally due to the
inhomogeneity of the distribution of the stars and not to the density
$n$). In other words two systems
having the same $n$ will have
their stars
slowed down differently according to the value of $p$. This is strictly
connected to the asymmetric origin of the dynamical friction. \\
In addition, by 
increasing $n$ the dynamical friction increases, just like in the 
homogeneous systems, but the increase is larger than the 
linear increase observed in homegeneous systems.

\section{Conclusion}

In this paper we calculated the distribution function 
and the first moment of $ {\bf f} = d {\bf F}/dt$ for an inhomogeneous system. 
We obtained an expression relating $ {\bf f} = d {\bf F}/dt$ to 
the degree of inhomogeneity in a gravitational system. 
In the last part of the paper
we showed the implications of this result on the dynamical friction in a
inhomogeneous system and in particular how
inhomogeneity acts as an amplifier of the asimmetry
effect giving rise to dynamical friction. 
Here we want to stress that the distribution function that we have 
obtained is valid for every inhomogeneous system and consequently 
it is more general than the distribution
function obtained by CN43, that is reobtained when we assume $p=0$ in our model.
Moreover Kandrup's (1980a) theory
of the stationary distribution $ W({\bf F})$ for inhomogeneous systems
is reobtained in the limit $ t \rightarrow \infty$. 

\begin{flushleft}
{\it Acknowledgements}
We are grateful to E. Recami and E. Spedicato for helpful and stimulating 
discussions during the period in which this work was performed and 
the referee Prof. P. Laguna for some useful comments and suggestions. 
\end{flushleft}

\appendix
\begin{flushleft}
{\bf Appendix: Derivation of $A({\bf k}, {\bf \Sigma})$}
\end{flushleft}
\renewcommand{\theequation}{A\arabic{equation}}
In order to find $A({\bf k}, {\bf \Sigma})$ we have to calculate
the integral in Eq. (\ref{eq:ot}) (see Sect. ~2).
We change the integration variable in Eq. (\ref{eq:ot}) from ${\bf r}$
to ${\bf \Phi}$
using the following relations:
\begin{equation}
d{\bf r} = - \frac{1}{2} (GM)^{1.5} {\Phi}^{- 4.5} d{\bf \Phi}
\label{eq:dod}
\end{equation}
\begin{equation}
r = \sqrt{\frac{GM}{\Phi}}
\label{eq:tred}
\end{equation}
we have that Eq. (\ref{eq:ot}) can be rewritten as:
\begin{eqnarray}
C({\bf k}, {\bf \Sigma}) & = & \frac{1}{2} \int_{0}^{\infty} \int_{- \infty}^{\infty} (G M)^{1.5}  d{\bf V} dM \nonumber \\
                            &   & \int_{- \infty}^{\infty} \tau \left[ 1 - e^{i({\bf k}{\bf \Phi} + {\bf \Sigma}{\bf \Psi})} \right] a \left( \frac{G M}{ \Phi} \right)^{-\frac{p}{2}} \Phi^{-\frac{9}{2}} d{\bf \Phi}
\label{eq:quat}
\end{eqnarray}
and if we express the product $ \Sigma \cdot \Psi$
as a function of $ \Phi$ (CN43 - Eq. 18)
we can
write $ C({\bf k}, {\bf \Sigma})$ in the following form: 
\begin{equation}
C({\bf k}, {\bf \Sigma}) = \frac{1}{2} G^{1.5} \int_{0}^{\infty} \int_{- \infty}^{\infty} M^{1.5}  D({\bf k}, {\bf \Sigma})  d{\bf V} dM
\label{eq:qui}
\end{equation}
where $ D({\bf k}, {\bf \Sigma})$ is given by:
\begin{equation}
D({\bf k}, {\bf \Sigma}) =  \int_{- \infty}^{\infty} \tau \left[ 1 - e^{i({\bf k}{\bf \Phi} + {\bf \Sigma}{\bf \Psi})} \right] \Phi^{- 4.5} d{\bf \Phi} \nonumber
\end{equation}
or equivalently:
\begin{eqnarray}
D({\bf k}, {\bf \Sigma}) & = & \int_{- \infty}^{\infty} \tau \left[ 1 - e^{(i{\bf k}{\bf \Phi})} \right] \Phi^{- 4.5} d{\bf \Phi} \; \nonumber \\
                            &   & +  \int_{- \infty}^{\infty} \tau [e^{i({\bf k}{\bf \Phi})}] \cdot [ 1 -  e^{(i{\bf \Sigma}{\bf \Psi})} ] \cdot \Phi^{- 4.5} d{\bf \Phi}
\label{eq:sed}
\end{eqnarray}
The first integral in Eq. (\ref{eq:sed}) can be easily calculated:
\begin{equation}
I_{1} =  \int_{- \infty}^{\infty} \tau \left[ 1 - e^{(i{\bf k}{\bf \Phi})} \right] \Phi^{- 4.5} d{\bf \Phi} = \frac{4 \pi a}{(G M)^{p/2}} \cdot B(p) \cdot k^{\frac{(3 - p)}{2}}
\label{eq:dic}
\end{equation}
with
\begin{equation}
B(p) = \int_{0}^{\infty} \frac{z - \sin{z}}{z^{(7 - p)/2}} \cdot d z
\end{equation}
For $ p = 0$  Eq. (\ref{eq:dic}) gives: 
\begin{equation}
I_{1} = \frac{8}{15} \cdot (2 \pi k)^{1.5} \nonumber
\end{equation}
that coincides with the first term in the right hand side of Eq. (22) 
in CN43.\\ 
The second integral is more difficult to evaluate and is given by: 
\begin{eqnarray}
I_{2} & = & \int_{- \infty}^{\infty} \tau [e^{(i{\bf k}{\bf \Phi})}]
\cdot [ 1 -  e^{(i{\bf \Sigma}{\bf \Psi})} ] \cdot \Phi^{- 4.5}
d{\bf \Phi} \nonumber \\
      &   & =  a \int_{- \infty}^{\infty}
      \left( \frac{\Phi}{G M} \right)^{p/2}
      [e^{(i{\bf k}{\bf \Phi})}] \; [ 1 -
      e^{(i{\bf \Sigma}{\bf \Psi})} ] \; \Phi^{-\frac{9}{2}} d {\bf \Phi}
\label{eq:dicio}
\end{eqnarray}
We have obtained this last result using Eq. (\ref{eq:nov}) and 
Eq. (\ref{eq:un}) (see Sect. ~ 2). Since we are interested in the moments of
$ {\frac{d \bf F}{d t}}$ for a given $ {\bf F}$ we
need only the behaviour of the function  $  A({\bf k}, {\bf \Sigma})$ for 
$ | {\bf \Sigma} | \Rightarrow 0$. To obtain this we expand the term 
$ [ 1 - e^{(i {\bf \Sigma} {\bf \Psi})} ] $
in powers of $ {\bf \Sigma}$ and
$ {\bf \Psi}$ in Eq. (\ref{eq:dicio}) and then we have: 
\begin{equation}
I_{2} =  \frac{a}{(G M)^{p/2}} \int_{- \infty}^{\infty} 
\left[ -i {\bf \Sigma}{\bf \Psi} + \frac{({\bf \Sigma}{\bf \Psi})^{2}}{2} 
\right] \; \Phi^{\frac{p-9}{2}} \; [e^{(i{\bf k}{\bf \Phi})}] d {\bf \Phi}  
\label{eq:dicia}
\end{equation}
or 
\begin{eqnarray}
I_{2} & = & \frac{- i a}{(G M)^{p/2}} \cdot  \int_{- \infty}^{\infty} 
{\bf \Sigma}{\bf \Psi} \cdot \Phi^{(p-9)/2} \cdot
[e^{(i{\bf k}{\bf \Phi})}]
d {\bf \Phi} +  \nonumber \\
      &   & \frac{a}{2 (G M)^{p/2}} \cdot  \int_{- \infty}^{\infty} 
({\bf \Sigma}{\bf \Psi})^{2} \cdot \Phi^{(p-9)/2} \cdot [e^{(i{\bf k}{\bf \Phi})}] d {\bf \Phi} \nonumber \\
      &   & = - D_1 + D_2
\label{eq:ven}
\end{eqnarray}
where $ {\bf \Sigma} \cdot {\bf \Psi}$ is given by
\begin{equation}
{\bf \Sigma} \cdot {\bf \Psi} = \Phi^{1.5} ({\bf \Sigma_{1}} {\bf V}) - 3 \Phi^{- 0.5} ({\bf \Phi}{\bf V}) \cdot ({\bf \Phi} {\bf \Sigma_{1}})
\label{eq:venuno}
\end{equation}
being ${\bf \Sigma_{1}}={\bf \Sigma}/(GM)^{0.5}$. 
Substituting these last equations in Eq. (\ref{eq:ven}) we have:
\begin{eqnarray}
D_{1}({\bf k}, {\bf \Sigma}) & = & \frac{ i a}{(G M)^{p/2}} \int_{- \infty}^{\infty} \left[ {\bf \Sigma_1}{\bf V} - 3 \cdot
\frac{({\bf \Phi}{\bf V}) \cdot ({\bf \Phi} 
{\bf \Sigma_1})}{\Phi^2} \right] \nonumber \\
                             &   &  \cdot \Phi^{(p-6)/2} \cdot [e^{(i{\bf k}{\bf \Phi})}] d {\bf \Phi} 
\label{eq:vendue}
\end{eqnarray}
Now, we evalutate this integral following CN43.
We introduce a system of coordinates with the $z$-axis in the direction of 
$ {\bf k}$ and letting 
$ {\bf \Sigma}_1 = (s_1, s_2, s_3)$, 
$ {\bf V} = (V_1, V_2, V_3)$ 
and $ {\bf l}_{\Phi}=(l, m, n) = (\sin{\theta} \cos{w}, \sin{\theta} \sin{w}, \cos{\theta})$.
The result of this integration is given by:
\begin{equation}
D_{1}({\bf k}, {\bf \Sigma}) = \frac{ 4 \pi i a}{(G M k)^{p/2}} \cdot (s_{1} V_1 + s_{2} V_2 - 2 s_{3} V_3) \cdot A(p)  
\label{eq:ventre}
\end{equation}
where
\begin{equation}
A(p) =  \int_{0}^{\infty} \left[  \frac{\sin{x}}{x^{(4 - p)/2}} -  \frac{3 \sin{x}}{x^{(8 - p)/2}} + \frac{3 \cos{x}}{x^{(6 - p)/2}} \right] \cdot d x  
\label{eq:venqua}
\end{equation}
For $ p = 0$ Eq. (\ref{eq:ventre}) becomes
\begin{equation}
D_{1}({\bf k}, {\bf \Sigma}) = -\frac{ 4 \pi i }{3} \cdot (s_{1} V_1 + s_{2} V_2 - 2 s_{3} V_3) 
\label{eq:vencin}
\end{equation}
which coincides with Eq. (37) in CN43.\\
The second integral in Eq. (\ref{eq:ven}) is:
\begin{equation}
D_{2}({\bf k}, {\bf \Sigma}) = \frac{ a}{2 (G M)^{p/2}} \cdot
\int_{- \infty}^{\infty} ({\bf \Sigma}{\bf \Psi})^{2}
\; \Phi^{\frac{p-9}{2}} \cdot [e^{(i{\bf k}{\bf \Phi})}] d {\bf \Phi}
\label{eq:vensei}
\end{equation}
If we define the variable
\begin{equation}
z = | {\bf k} | \cdot | {\bf \Phi} |
\label{eq:venset}
\end{equation}
and we introduce Eq. (\ref{eq:venuno}) in Eq. (\ref{eq:vensei}) we obtain:
\begin{eqnarray}
D_{2}({\bf k}, {\bf \Sigma}) & = & \frac{ a}{2 (G M)^{p/2}} \; k^{- \frac{3+p}{2}} \int_{0}^{\infty} \int_{- 1}^{1} \int_{0}^{2 \pi} [ e^{(i z t)}] \cdot \nonumber \\
                             &   & \left[ {\bf \Sigma}_1 {\bf V} - 3 
( {\bf l}_{\Phi} \cdot {\bf V})( {\bf l}_{\Phi} \cdot {\bf \Sigma}_1) \right]^{2} \, z^{\frac{p + 1}{2}} dw dt dz
\label{eq:venot}
\end{eqnarray}
where $ t = cos \theta$ and $ w$ is the azimuthal angle.
The integral can be calculated regarding $ z$ and $ t$ as complex variables
and using an appropriate chosen contour (see Chandrasekhar \& von
Newmann 1942).
The result of the integration in terms of the original variable
${\bf \Sigma}=(\Sigma_1,\Sigma_2,\Sigma_3)$ is:
\begin{eqnarray}
D_{2}({\bf k}, {\bf \Sigma}) & = & \frac{ - \pi a k^{- (3 + p)/2}}
{2 (G M)^{p/2}} \, \Gamma \left( \frac{p}{2} +
\frac{3}{2} \right)  \left\{ \cos \left[ -
\frac{\pi}{4} \left( \frac{p}{3} + 1 \right) \right] \right. \nonumber \\
                             &   &  - \sin \left[ - \frac{\pi}{4} 
                                \left( \frac{p}{3} + 1 \right) \right]
                                \left\}  \, \{ \Sigma_{1}^{2} [
a_{\ast} V_{1}^{2} + b V_{2}^{2} + c V_{3}^{2} ] + \right. \nonumber \\ 
                             &   & \Sigma_{2}^{2} [ 
 b V_{1}^{2} + a_{\ast} V_{2}^{2} + c V_{3}^{2} ]  + \Sigma_{3}^{2} [ 
 c V_{1}^{2} + e V_{2}^{2} + d V_{3}^{2} ] +   \nonumber \\
                             &   & 
+ f (\Sigma_{2} \Sigma_{3} V_2 V_3 +
\Sigma_{1} \Sigma_{3} V_1 V_3) + e \Sigma_{2} \Sigma_{1} V_2 V_1 \}  
\label{eq:venno}
\end{eqnarray}
where
\begin{equation}
a_{\ast} =  \frac{- 2}{p + 1} - \frac{ 24}{(p + 1) (p - 3) } - \frac{ 216}{(p + 1) (p - 3) (p - 7) } \nonumber
\end{equation}
\begin{equation}
b =  \frac{- 72}{(p + 1) (p - 3) (p - 7) } \nonumber
\end{equation}
\begin{equation}
c =  \frac{- 36}{(p - 3) (p - 7) } 
\label{eq:tren}
\end{equation}
\begin{equation}
d =    \frac{- 2}{p + 1} -  \frac{12}{3 - p} + \frac{18}{7 - p}  \nonumber
\end{equation}
\begin{equation}
e =  \frac{- 4}{p + 1} - \frac{ 48}{(p + 1) (p - 3) } - \frac{ 288}{(p + 1) (p - 3) (p - 7) } \nonumber
\end{equation}
\begin{equation}
f = \frac{- 4}{p + 1} - \frac{ 24}{(p + 1) (p - 3) } - \frac{12}{3 - p} + \frac{144}{(p - 3) (p - 7) } \nonumber
\end{equation}
then we have obtained:
\begin{eqnarray}
D({\bf k}, {\bf \Sigma}) & = & I_{1} + I_{2} = \frac{ 4 \pi a k^{-
\frac{3+p}{2}}}{(G M)^{p/2}} \, B(p) \; - \; \frac{ - 4 \pi i a
k^{- p/2}}{(G M)^{p/2}} \cdot \nonumber \\
                         &    & (\Sigma_{1} V_1 +
                            \Sigma_{2} V_2 - 2 \Sigma_{3} V_3)
                            \cdot A(p) \; + \; D_2
\label{eq:trenuno}
\end{eqnarray}
Now substituting Eq. (\ref{eq:trenuno}) in Eq. (\ref{eq:qui}) we have:
\begin{eqnarray}
C({\bf k}, {\bf \Sigma}) & = & 2 \pi a (G \overline{M})^{(3 - p)/2} k^{(3 - p)/2} B(p) - 2 \pi i a k^{-p/2}  \nonumber \\ 
                         &   & \cdot  (G \overline{M})^{(2 - p)/2} (\Sigma_{1} V_1 + \Sigma_{2} V_2 - 2 \Sigma_{3} V_3) \cdot A(p)  + \nonumber \\
                         &   &  - \frac{\pi a}{2} (G \overline{M})^{(1 - p)/2} k^{- (3 + p)/2} \, \Gamma \left( \frac{p}{2} +  \frac{3}{2} \right) \nonumber \\
                         &   & \cdot \left\{ \cos{ \left[ - \frac{\pi}{4} 
\left( \frac{p}{3} + 1 \right) \right]} - \sin{ \left[ - \frac{\pi}{4} \cdot \left( \frac{p}{3} + 1 \right) \right]} \right\}  \nonumber \\ 
                         &   & \cdot \{ \Sigma_{1}^{2} [ 
a \overline{V}_{1}^{2} + b \overline{V}_{2}^{2} + c \overline{V}_{3}^{2} ] + \Sigma_{2}^{2} [ 
b \overline{V}_{1}^{2} + a \overline{V}_{2}^{2} + c \overline{V}_{3}^{2} ] + \nonumber \\ 
                         &   &  + \Sigma_{3}^{2} [ 
c \overline{V}_{1}^{2} + e \overline{V}_{2}^{2} + d \overline{V}_{3}^{2} ] +  f (\Sigma_{2} \Sigma_{3} \overline{V_2 V_3} +  \nonumber \\
                         &   & + \Sigma_{1} \Sigma_{3} \overline{V_1 V_3}) + e \Sigma_{2} \Sigma_{1} \overline{V_2 V_1} ] \}
\label{eq:trendue}
\end{eqnarray}
where we have used bars to indicate that the corresponding
quantities have been averaged with the weight function $\tau({\bf V},M)$. \\
If the distribution of the velocities of the field stars, $ {\bf u}$, is spherical and the test star moves with velocity $ {\bf v}$ we have: 
\begin{equation}
{\overline {\bf V}} =  {\overline {\bf u}} - {\overline {\bf v}} = - {\bf v}  \nonumber
\end{equation}
and also:
\begin{equation}
{\overline  V_{u}} =  {\overline v_{u}}   \nonumber
\end{equation}
\begin{equation}
{\overline  V_{u}^{2}} =  \frac{1}{3} | {\bf u}|^{2} +  v_{u}^{2} 
\label{eq:trentre}
\end{equation}
\begin{equation}
{\overline  V_{u} V_{\nu}} =   v_{u} v_{\nu}  \nonumber
\end{equation}
and if we use the system of coordinates introduced by CN43 
that is  $ v_1 = | {\bf v} | sen{\gamma}$, 
$ v_2 = 0 $ and $ v_3 = | {\bf v} | cos{\gamma}$ where $ \gamma$ is 
the angle between $ {\bf k}$ and $ {\bf v}$ we can simplify Eq. (\ref{eq:trendue}) and then 
the Eq. (\ref{eq:setbis}) (see Sect. ~2) becomes:
\begin{eqnarray}
A({\bf k}, {\bf \Sigma}) & = & e^{- n C({\bf k}, {\bf \Sigma})} =  
e^{ [- \; \;  \frac{\alpha}{2} ( G M k )^{\frac{3 - p}{2}} \cdot 
B(p) \; + \; \frac{i \alpha}{2} k^{\frac{- p}{2}} } \cdot  \nonumber \\
                         &   &  ( G M )^{\frac{2 - p}{2}} \left| {\bf v} \right| \cdot [ \Sigma_{1}  \sin{\gamma} \; - \; 2 \Sigma_{3} \cos{\gamma} ] \cdot A(p) + \nonumber \\
                         &   &  - \; \frac{\alpha}{8} ( G M )^{\frac{1 - p}{2}} k^{\frac{ -(3 + p)}{2}} \cdot  \Gamma \left( \frac{p}{2} + \frac{3}{2} \right)  
\nonumber \\
                         &   & \cdot \left \{ \cos{\left[ - \frac{\pi}{4} \left( \frac{p}{3} + 1 \right) \right]} \; - \; \sin{\left[ - \frac{\pi}{4} \cdot \left( \frac{p}{3} + 1 \right) \right]}  \right
\} \cdot \nonumber \\ 
                         &   & \left \{ \left| {\bf u} \right|^{2} \cdot \left[ \frac{1}{3} \; ( a + b + c ) ( \Sigma_{1}^{2} +  \Sigma_{2}^{2} ) \; + \;  \frac{\Sigma_{3}^{2}}{3} 
( 2 c + d ) \right] \right. \nonumber \\
                         &   & + \; \left| {\bf v} \right|^{2} \cdot \left[ (a \sin^{2}{\gamma} + c \cos^{2}{\gamma}) \Sigma_{1}^{2} \; + \; (b \sin^{2}{\gamma} +  \right. \nonumber \\
                         &   & + \; c \cos^{2}{\gamma}) \Sigma_{2}^{2} + \; ( c \sin^{2}{\gamma} + d \cos^{2}{\gamma}) \Sigma_{3}^{2} \; + \nonumber \\
                         &   &
+ \; f \Sigma_{1} \Sigma_{3} 
\sin{\gamma} \cos{\gamma} \left] \left\} \right. \right.]
\label{eq:trenci}
\end{eqnarray}
where $ \alpha = 4 \pi n a $. We define, now, the following constants and 
functions :
\begin{equation}
\tilde{a} =  \frac{\alpha}{2} ( G M )^{\frac{3 - p}{2}} \cdot B(p) \nonumber
\end{equation}
\begin{equation}
g =   \frac{\alpha}{2} ( G M )^{\frac{2 - p}{2}}
\cdot  | {\bf v} | \cdot A(p) \nonumber
\label{eq:cas}
\end{equation}
\begin{equation}
p({\bf k}, {\bf \Sigma}) =  k^{\frac{-p}{2}}
\cdot [ \Sigma_1 \sin{\gamma} \; - \; 2 \Sigma_3 \cos{\gamma} ] \nonumber
\label{eq:cass}
\end{equation}
\begin{eqnarray}
\tilde{b} & = & \frac{\alpha}{8} ( G M )^{\frac{1 - p}{2}} \cdot \Gamma \left( \frac{p}{2} +  \frac{3}{2} \right) \; \cdot \nonumber \\ 
          &   & \left\{ \cos{ \left[ - \frac{\pi}{4} \left( \frac{p}{3} + 1 \right) \right]} - \sin{ \left[ - \frac{\pi}{4} \cdot \left( \frac{p}{3} + 1 \right) \right] } \right\} \cdot  | {\bf u} |^{2}  
\end{eqnarray}
\begin{equation}
Q({\bf \Sigma}) =  \left[ \frac{1}{3} 
\; ( a_{\ast} + b + c ) ( \Sigma_{1}^{2} +  \Sigma_{2}^{2} )+
\frac{\Sigma_{3}^{2}}{3}
( 2 c + d ) \right] \nonumber
\end{equation}
\begin{eqnarray}
R({\bf \Sigma}) & = & (a_{\ast} \sin^{2}{\gamma} + c \cos^{2}{\gamma}) \Sigma_{1}^{2} \; + \; (b \sin^{2}{\gamma} + c \cos^{2}{\gamma}) \Sigma_{2}^{2} \nonumber \\
                &   & + \; ( c \sin^{2}{\gamma} + d \cos^{2}{\gamma}) \Sigma_{3}^{2} \; + \; f \Sigma_{1} \Sigma_{3} 
\sin{\gamma} \cos{\gamma} \nonumber
\end{eqnarray}
\begin{equation}
k =  \frac{| {\bf v} |^{2}}{| {\bf u} |^{2}} \nonumber
\end{equation}
so we can re-write Eq. (\ref{eq:trenci}) as:
\begin{eqnarray}
A({\bf k}, {\bf \Sigma}) & = & e^{  -\tilde{a} k^{\frac{3-p}{2}}} \{1- \; i g p({\bf k}, {\bf \Sigma})   \nonumber \\
                            &   & + \; \tilde{b} k^{\frac{- (3 
+ p)}{2}} ] \cdot  [ Q({\bf \Sigma}) + k R({\bf \Sigma}) ] \} 
\label{eq:trense}
\end{eqnarray} 


\begin{thebibliography}{}
\bibitem {ah1} Ahmad A., Cohen L., 1973, ApJ 179, 885
\bibitem {ah2} Ahmad A., Cohen L., 1974, ApJ 188, 469 
\bibitem{} Antonuccio-Delogu V., Colafrancesco S., 1994, ApJ 427, 72   
\bibitem{bas} Bahcall N.A., Soneira R.M., 1983, ApJ 262, 419 
\bibitem{} Berkenstein J.D., Maoz E., ApJ 1992, 390, 79
\bibitem{} Binney J., Tremaine S., 1987, " Galactic Dynamics " (Priceton: Princeton University Press)
\bibitem{} Bontekoe, T. R., van Albada, T. S., 1987, MNRAS, 224, 349
\bibitem{} Chandrasekhar S., 1941, ApJ 94, 511
\bibitem{} Chandrasekhar S., 1943a, Rev. Mod. Phys. 15, 1
\bibitem{} Chandrasekhar S., 1943b, ApJ 97, 255
\bibitem{} Chandrasekhar S., 1943c, ApJ 97, 263
\bibitem{} Chandrasekhar S., 1943d, ApJ 98, 25
\bibitem{} Chandrasekhar S., 1943e, ApJ 98, 47
\bibitem{cha2} Chandrasekhar S., 1944a, ApJ 99, 47
\bibitem{cha3} Chandrasekhar S., 1944b, ApJ 99, 25
\bibitem{} Chandrasekhar S., von Neumann J., 1942, ApJ 95, 489
\bibitem{} Chandrasekhar S., von Neumann J., 1943, ApJ 97, 1 (CN43)
\bibitem{} Del Popolo A., 1994, Ph.D. dissertation, Istituto di 
Astronomia dell' Universit\`a di Catania 
\bibitem{} Del Popolo A., Gambera M., 1996, A\&A 308, 373
\bibitem{} Del Popolo A., Gambera M., 1997, A\&A 321, 691
\bibitem{del} Del Popolo A., Gambera M., Antonuccio-Delogu, V., 1996,  Mem. So
c. Astr. It. 67, n. 4, 957
\bibitem{} Dominguez-Tenreiro R., Gomez-Flechoso M. A., 1998,
{\it preprint SISSA astro-ph/9709290}
\bibitem{} Elson R., Hut P., Inagaki S., 1987, ARA\&A 25, 565
\bibitem{gam1} Gambera M., 1997, Ph.D. dissertation, Istituto di 
Astronomia dell' Universit\`a di Catania
\bibitem{} Holtsmark P.J., 1919, Phys. Z. 20, 162
\bibitem{} Kashlinsky A., 1986, ApJ 306, 374
\bibitem{} Kashlinsky A., 1987, ApJ 312, 497
\bibitem{kan} Kandrup H.E., 1980a, Phys. Rep. 63, n. 1, 1
\bibitem{kdr} Kandrup H.E., 1980b, ApJ 244, 1039
\bibitem{kdr} Kandrup H.E., 1983, Ap.\&S.S. 97, 435
\bibitem{} Maoz E., 1993, MNRAS 263, 75
\bibitem{lid} Liddle A.R., Lyth D.H., 1993, Phys. Rep. 231, n 1, 2 
\bibitem{pee} Peebles P.J.E., 1980, "The large scale structure of the Universe", (Priceton: Princeton University Press)
\bibitem{} Rosenblunth M.N., Mc Donald W.M., Judd D.L., 1957, Phy. Rev. 107, 1
\bibitem{sar} Sarazin C., 1988, " X-ray emission from Clusters of Galaxies", (Cambridge: Cambridge Univ. Press)
\bibitem{} Seguin P., Dupraz C., 1996, A\&A 310, 757
\bibitem{stw} Strauss M.A., Willick J.A., 1995, Phys. Rept. 261, 271
\bibitem{} Tremaine S., Weinberg M.D., 1984, MNRAS 209, 729
\bibitem{} Weinberg M.D., 1986, ApJ 300, 93
\bibitem{} White S.D.M., 1976, MNRAS 174, 19
\bibitem{whi} White S.D.M., Briel U.G., Henry J.P., 1993, MNRAS 261, L8
\bibitem{} Wybo M., Dejonghe H., 1996, accepted A\&A {\it preprint SISSA astro-ph/9502104}
\bibitem{} Zwart S.F.P., Tout C.A., Lee, H.M., to appear in Highlights of Astronomy Vol. 11, Kluwer Academic Publishers, ed. Johannes Andersen 
{\it preprint SISSA astro-ph/9710209}
\end{thebibliography}
\end{document}